\documentclass[aps,twocolumn,showpacs,superscriptaddress]{revtex4-1}
\usepackage{amsmath}
\usepackage{amssymb}
\usepackage{bbm}
\usepackage{graphicx}
\usepackage[colorlinks=true,citecolor=blue,linkcolor=red,linktocpage=true,pagebackref=false]{hyperref}
\usepackage{color}
\usepackage{varwidth}
\usepackage{times}
\usepackage{textcomp} 
\usepackage{mathtools}
\mathtoolsset{centercolon}

\newcommand{\be}{\begin{eqnarray}}
\newcommand{\ee}{\end{eqnarray}}
\newcommand{\ben}{\begin{eqnarray*}}
\newcommand{\een}{\end{eqnarray*}}
\newcommand{\bee}{\begin{equation}}
\newcommand{\eee}{\end{equation}}
\newcommand{\one}{\mathbbm{1}}
\newcommand{\Tr}[1]{\mathrm{Tr}\left[ {#1} \right]}

\newcommand{\ket}[1]{\left|{#1}\right\rangle}
\newcommand{\bra}[1]{\left\langle{#1}\right|}

\definecolor{adrian}{rgb}{0,0,0}
\definecolor{adrian2}{rgb}{0,0,0}
\definecolor{comment}{rgb}{0,0,0}
\definecolor{geraldine}{rgb}{0,0,0}
\definecolor{jens}{rgb}{0,0,0}
\newcommand{\ad}[1]{{\color{adrian} #1}}

\newcommand{\ger}[1]{{\color{geraldine} #1}}
\newcommand{\je}[1]{{\color{jens} #1}}

\begin{document}

\title{Continuous matrix product state tomography of quantum transport experiments}

\author{G.\ Haack} 
\affiliation{CEA Grenoble, 17 rue des Martyrs, 38000 Grenoble, France}
\affiliation{Dahlem Center for Quantum Complex Systems and Fachbereich Physik, Freie Universit\"at Berlin, 14195 Berlin, Germany}

\author{A.\ Steffens}
\affiliation{Dahlem Center for Quantum Complex Systems and Fachbereich Physik, Freie Universit\"at Berlin, 14195 Berlin, Germany}
\author{J.\ Eisert}
\affiliation{Dahlem Center for Quantum Complex Systems and Fachbereich Physik, Freie Universit\"at Berlin, 14195 Berlin, Germany}
\author{R.\ H\"ubener}
\affiliation{Dahlem Center for Quantum Complex Systems and Fachbereich Physik, Freie Universit\"at Berlin, 14195 Berlin, Germany}
\date{\today}

\begin{abstract}
In recent years, a close connection between the description of open quantum systems, the input-output formalism of quantum optics, and continuous matrix product states in quantum field theory has been established. So far, however, this connection has not been extended to the condensed-matter context. In this work, we substantially develop further and apply a machinery of continuous matrix product states (cMPS) to perform tomography of transport experiments. We first present an extension of the tomographic possibilities of cMPS by showing that reconstruction schemes do not need to be based on low-order correlation functions only, but also on low-order counting probabilities. We show that fermionic quantum transport settings can be formulated within the cMPS framework. This allows us to present a reconstruction scheme based on the measurement of low-order correlation functions that provides access to quantities that are not directly measurable with present technology. Emblematic examples are high-order correlations functions and waiting times distributions (WTD). The latter are of particular interest since they offer insights into short-time scale physics. We demonstrate the functioning of the method with actual data, opening up the way to accessing WTD within the quantum regime.
\end{abstract}

\maketitle



\section{Introduction}
\label{intro}
%
%

\emph{Continuous matrix product states} (cMPS) have recently been recognised as powerful and versatile
descriptions of certain one-dimensional quantum field states \cite{cMPS1, cMPS2,cMPSCalculus}. As continuum limits of the matrix product states (MPS)---\linebreak a well-established type of tensor network states underlying the density-matrix renormalisation group machinery~\cite{Schollw2}---they introduce the intuition developed in quantum lattice models to the realm of quantum fields, offering similar conceptual and numerical tools. In the cMPS framework, interacting quantum fields such as those described by Lieb-Liniger models have been studied, both in theory \cite{cMPS1,Draxler,Quijandria14} and in the context of experiments with ultra-cold atoms \cite{cMPSTomographyShort}.

On a formal level, continuous matrix product states are intricately related to Markovian \emph{open quantum systems} \cite{cMPS1, cMPS2}: The open quantum system takes the role of an ancillary system in a sequential preparation picture of cMPS. Elaborating on this formal analogy, cMPS can capture properties of fields that are coupled to a finite dimensional open quantum system. This connection has been fleshed out already in the description of light emitted from cavities in cavity-QED \cite{cMPS2,CavityQED} in the quantum optical context, under the keyword of the \emph{input-output formalism} \cite{Gardiner}.

Another methodological ingredient to this work is that cMPS have been identified as tools to perform efficient \emph{quantum state tomography} of quantum field systems \cite{cMPSTomographyShort,LongPaper,Wick,Bruderer13,Guta}, related to other approaches of tensor network quantum tomography \cite{MPSTomo,Wick}.
These efforts are in line with the emerging mindset that for quantum many-body and quantum field states, tomography and state reconstruction only make sense within a certain statistical model or a variational class of states. Importantly, in our context at hand, it turns out that cMPS can be reconstructed from the knowledge of low-order correlation functions alone. This is a very attractive feature of cMPS: Recently, a reconstruction scheme has successfully been applied to data on quantum fields obtained with ultra-cold Bose gases~\cite{cMPSTomographyShort}. Read in the mindset of open quantum systems, cMPS tomography can be interpreted as open system tomography by monitoring the environment of the open quantum system (see Section~\ref{sec:cmps_tomo}).

\begin{figure*}
\includegraphics[width=0.7\textwidth]{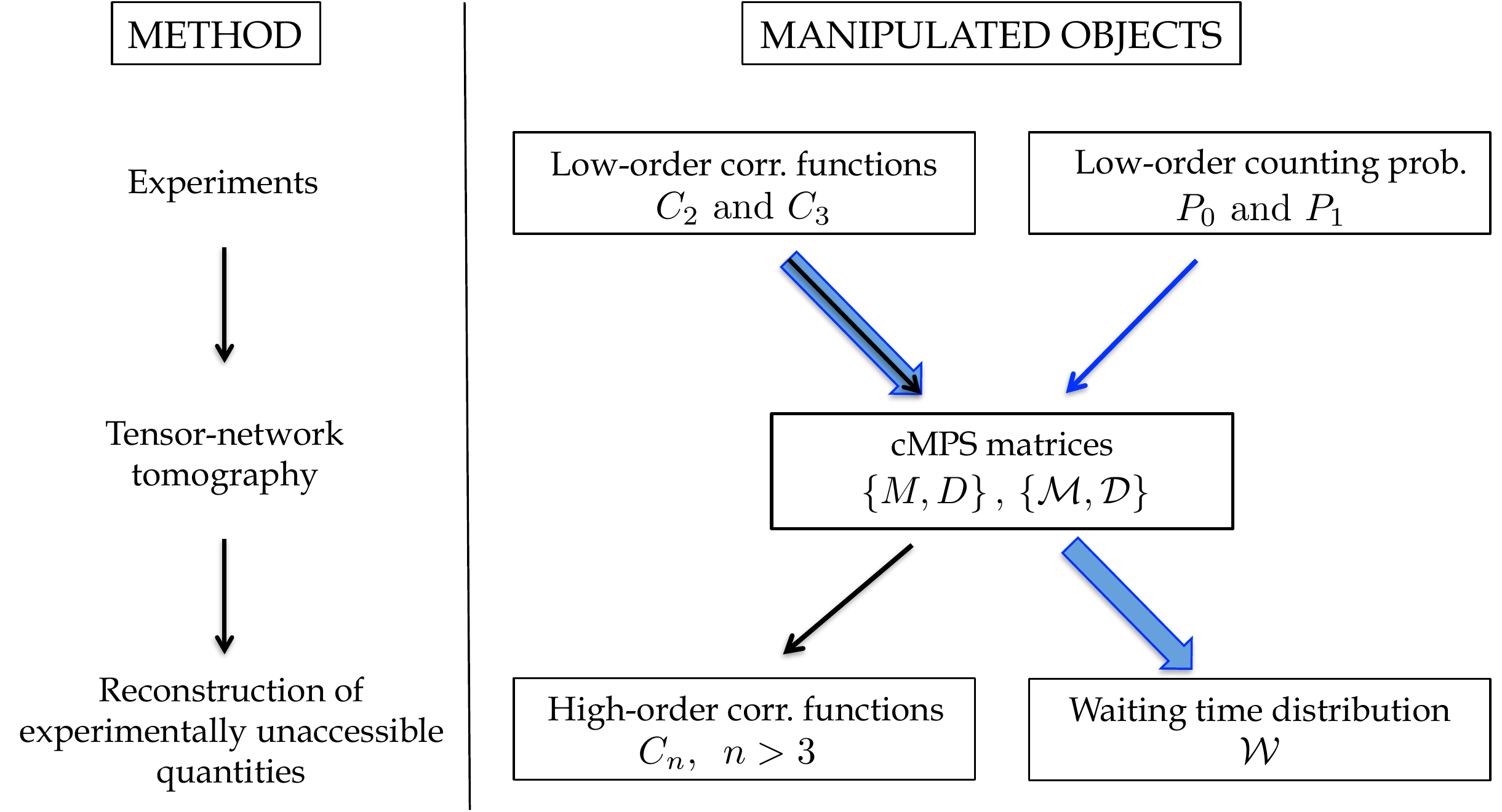} 
\caption{Extension and applications of cMPS based tomography (see Sections~\ref{sec:proof}--\ref{sec:WTD}). Previous works \cite{LongPaper,cMPSTomographyShort} have shown that measurements of low-order correlation functions $C_n$ with $n=2,3$ are sufficient to access higher-order correlations using cMPS tomography based on the cMPS matrices $M$ and $D$ (right side, black arrows). The first achievement of this work is the formal demonstration that measurements of low-order counting probabilities  constitute an alternative to measurements of low-order correlation functions for carrying out cMPS tomography (blue arrow). We show that the probabilities to detect zero or one particle ($P_0$ and $P_1$ respectively) are sufficient to reconstruct alternative cMPS parameter matrices $\mathcal{M}$ and $\mathcal{D}$, from which higher-order correlation functions can be computed. The second achievement of this work is to extend the applicability of this cMPS machinery to quantum transport experiments. As a main illustration, we show that cMPS-based tomography provides an access to the distribution of waiting times $\mathcal{W}$. The according statistics are not directly measurable due to experimental limitations on single-particle detectors. However, we demonstrate with experimental data that they can in fact be reconstructed from the knowledge of low-order correlation functions (broad blue arrows).
}
\label{fig:relations}
\end{figure*}

In this work, these methodological components will be put into a different physical context and substantially developed further as illustrated in Fig.~\ref{fig:relations}. At the heart of the analysis is a tomographic approach, applied to an open quantum system, yet brought to a new level. In Section~\ref{sec:proof}, we extend the set of tomographic methods within the cMPS framework, showing that the dynamics of the ancillary system and of the whole open quantum system is not only accessible from low-order correlation functions, but also from low-order counting statistics. Specifically, we prove that for generic systems, the two density functions $P_0$ and $P_1$---which express the probability of detecting zero and one particle, respectively, as a function of the time since the last detection---provide sufficient knowledge to successfully perform tomography of the open quantum system.
The physical application of the established methods will also be different from the cavity-QED or the quantum field context: Here, we treat \emph{fermionic quantum transport experiments} within the cMPS framework. 

In a general transport setting, a scatterer is coupled to a left reservoir (the ``source'') and a right reservoir (the ``drain''). Fermions (with or without a spin degree of freedom) can be seen as jumping in and out of the scattering region from the source to the drain and can be described by a leaking-out fermionic quantum field. In Section~\ref{sec:QT}, we show how the dynamics of the open quantum system (scatterer and leaking-out fermionic field) can be encoded into a cMPS state vector. 
To provide the reader with an intuition about the equivalence between the cMPS language and a more traditional Hamiltonian formulation, we will consider one of the simplest setups in quantum transport: a single-level quantum dot weakly coupled to two reservoirs. These results are also valid for transport experiments of ultra-cold fermions between a ``hot'' and a ``cold" reservoir as recently realised in Refs.~\cite{Brantut12, Brantut13}.

This will clear the way for making use of the tomographic possibilities offered by the cMPS formalism to access various quantities in quantum transport that are not yet measurable with current experimental technologies (see Fig.~\ref{fig:relations}). Emblematic examples are higher-order charge correlation functions and the distribution of waiting times (WTD, see Section~\ref{sec:WTD}).

The waiting time is defined as the time interval between the arrivals of two \textit{consecutive} electrons. Therefore, the WTD provides a privileged access to short-time physics, short-range interactions and the statistics of the particles. As such, it has gained a lot of attention recently \cite{Brandes08, Albert11, Albert12, Thomas13, Rajabi13, Dasenbrook14, Albert14, Tang14}, but WTDs suffer from their difficulty to be measured effectively: Measuring WTDs requires the detection of single events while ensuring that no events have been missed---for instance, due to the dead time of the detector.

With present technologies, WTDs in transport experiments can be measured when the injection rate of electrons is within the kHz range. Here, the current trace is resolved in time and the WTD can be directly deduced from it. This is for instance the case in the experiments ~\cite{Gustavsson06, Ubbelohde12}. It is important to keep in mind that experiments in the kHZ range can only mimic transport properties of classical particles---no quantum effects such as the statistics of the injected particles or the coherence properties of the scatterer can be observed at these frequencies. To observe quantum mechanical effects within the setup, one needs to move to the GHz regime, which can be achieved either with DC sources with a typical bias of tens of
meV, or with periodically driven sources at GHz frequencies \cite{Feve07, Blumenthal07, Fujiwara08, Maire08, Dubois13}. In the GHz range, the current trace can not be resolved in time so that the measurement of the WTD is not feasible at present. In contrast, second- and third-order correlation functions have been proven to be feasible, 
but are accompanied by exceedingly difficult measurement prescriptions
\cite{Dubois13, Gabelli13}.

In the light of this discussion, we propose in Section~\ref{sec:WTD} an indirect way to access the WTD with methods that are within reach of the experimental state of the art. Namely, the dynamics of the full open quantum system is accessed from  measurements of low-order correlation functions (typically second- or third-order). This is made possible with a cMPS formulation of the transport experiments as explained in the following section.

We illustrate this indirect path of accessing the WTD by considering real data obtained in the experiment of Ref.~\cite{Ubbelohde12} where single electrons tunnel through a single-level quantum dot in the kHz regime. Both the current trace resolved in time and the two- and three-point correlation functions have been measured. Although no quantum effects are present as explained above, the data allows us to demonstrate a very good agreement between the WTD deduced directly from the current trace and the WTD obtained via our reconstruction scheme based on the data of the correlation functions. This gives substance to our protocol based on cMPS to access the WTD with present technologies. We claim that this method remains valid in the GHz frequency range and for more complex systems such as a double quantum-dot coupled to two reservoirs---which would exhibit quantum coherence effects---and for quantum transport experiments with fermionic quantum gases.



\section{cMPS tomography}
\label{sec:cmps_tomo}

In order to present a self-contained analysis, we start by reviewing the cMPS formulation of capturing a finite dimensional open quantum system~\cite{cMPS2} and the tomography procedure of reconstructing the relevant cMPS parameter matrices~\cite{LongPaper}. Consider an open quantum system (in cMPS terms the \emph{ancillary system}) with dimension $d$ (called \emph{bond dimension} in that context) and interacting with one or more quantum fields that are described by field operators $\hat{\psi}_\alpha$ for different fields $\alpha$. Its dynamics can in general be represented by different mathematical objects:

\begin{itemize}

\item[a)] The master equation in Lindblad form, which governs the evolution of the ancillary system described by its state vector $\ket{\Psi}$ defined on the Hilbert space $\mathcal{H}$ of dimension $d\times d$. The degrees of freedom of the coupled fields are traced out in this approach. 

\item[b)] The set of $n$-point correlation functions of the coupled fields. 


\item[c)] The full counting statistics of the field system, i.e. the complete set of cumulants of the probability distribution of transferred particles. The $n$-th cumulant of the generating function is linked to the $n$ moments of this distribution, which correspond to the $n$-point correlation function.

\item[d)] The cMPS state vector $\ket{\psi_\text{cMPS}}$, which we now introduce.
\end{itemize}

\subsection{cMPS reconstruction from correlation functions}
\label{subsec:recon}

An intuitive way of establishing the cMPS state vector $\vert \psi_\text{cMPS} \rangle$ consists in starting from the well-known Lindblad equation. This equation describes the evolution of the state $\rho$ in time via the Liouvillian superoperator $\mathcal{L}$,
\be
\label{eq:lindblad}
\begin{split}
\dot{\rho} &= \mathcal{L} [\rho] \\
&= -\frac{\mathrm{i}}{\hbar} [K, \rho] - \frac{1}{2} \sum_{\alpha=1}^p \left(\left\{ R_\alpha^\dagger R_\alpha, \rho \right\} - 2 R_\alpha \rho R_\alpha^\dagger\right).
\end{split}
\ee
The first term relates to the free evolution via a Hamiltonian $K\in \mathbb{C}^{d\times d}$, while the last two terms describe the coupling to the environment (the according operator is known as the dissipator). The matrices $R_\alpha\in \mathbb{C}^{d\times d}$, $\alpha=1,\dots,p$, correspond to jump operators between the system and external quantum fields $\{\hat{\psi}_\alpha\}$.  The matrices $K$ and $\{R_\alpha\}$ completely characterise the evolution of the system.

Making use of the Choi-Jamio\l{}kowski isomorphism \cite{Jamiolkowski} (which maps linear superoperators from $\mathcal{H}_1$ to $\mathcal{H}_2$ to linear operators acting on $\mathcal{H}_1 \otimes \mathcal{H}_2$) the state $\rho$ is mapped to a state vector $\ket{\rho}$ and the Liouvillian  $\mathcal{L}$ to the matrix $T$  
\cite{cMPS1, cMPS2} with
\be
\label{eq:Liouvillian}
T = Q^* \otimes  \one + \one \otimes Q + \sum_\alpha R_\alpha^* \otimes R_\alpha.
\ee
The matrix $T\in\mathbb{C}^{d^2\times d^2}$ is known as the \emph{transfer matrix} and the matrix $Q$ is defined as
\be
\label{eq:Q}
Q = -\mathrm{i} K - \frac{1}{2} \sum_\alpha R_\alpha^\dagger R_\alpha.
\ee
Formally, the isomorphism introduced above is defined by the following relations for an operator and the product of operators
\be
\label{eq:def_iso}
\begin{split}
\rho &\mapsto \ket{\rho}\\
A^\dag \rho B &\mapsto ( A^* \otimes B ) \ket{\rho}.
\end{split}
\ee
Being closely connected to $K$ and $\{R_\alpha\}$ introduced above, the knowledge of the matrix and $T$ and its components provides access to the dynamics of the open quantum system, and allows to directly derive the according Lindblad equation.

The (translationally invariant) cMPS state vector $\ket{\Psi_\text{cMPS}}$ on the interval $[0,L]$ is defined in terms of the matrices $Q, \{ R_\alpha\}$ and the field operators $\hat{\psi}^\dagger_\alpha$ by 
\begin{align}
\label{eq:def_cmps}
\ket{\psi_\text{cMPS}} =  \mathrm{Tr}_{\text{anc}}
 \bigg[\mathcal{P}\hspace{-.05em} \exp\hspace{-.2em} \int_0^L\hspace{-.6em} \mathrm{d}x  \Big( Q \otimes \hat{\one} + \sum_\alpha R_\alpha \otimes \hat{\psi}_\alpha^\dagger(x)\Big)\bigg] \vert \Omega \rangle.
\end{align}
This expression is related to the path ordered exponential that arises when integrating the Lindblad equation. The embedding of the cMPS state vector $\vert \psi_\text{cMPS} \rangle$ into Fock space becomes clear when expanding the path ordered exponential $\mathcal{P}\exp$. For more details, we refer to Ref.~\cite{cMPSCalculus} where the authors  formulate the cMPS in different representations such as the Fock space and a path integral formulation. After integration, the ancillary system is traced out via $\mathrm{Tr}_{\text{anc}}$ and the resulting term is applied to the vacuum state vector $\vert \Omega \rangle$ where  $\hat{\psi}_\alpha \vert \Omega \rangle =0$ for each $\alpha$. 

Compared to the Lindblad equation, the main difference is that the degrees of freedom of the ancillary system are traced out such that its dynamics is mapped into the dynamics of the coupled quantum fields $\{ \hat{\psi}_\alpha\}$. The evaluation of expectation values of field operators leads to expressions that only contain quantities from the ancillary system, and information about the ancillary system can be inferred from according field operator measurements. For the sake of clarity, we restrict ourselves to the case where a single coupled quantum field, denoted as $\hat{\psi}_\beta$, is measured. 

The density-like correlation functions of the measured quantum field $\hat{\psi}_\beta$ then read
\be
\label{eq:def_corr}
C_{n}({\bf x}) = \bra{\psi_{\text{cMPS}}} \hat n(x_1) \dots \hat n(x_n) \ket{\psi_{\text{cMPS}}},
\ee
where ${\bf x} := (x_1, \ldots, x_n)$ 
and $\hat n :=  \hat{\psi}_\beta^\dag \hat{\psi}_\beta$. 
According to the calculus of expectation values in the cMPS setting~\cite{cMPSCalculus}, inserting Eq.\ \eqref{eq:def_cmps} into Eq.~\eqref{eq:def_corr} in the thermodynamic limit $L\rightarrow\infty$ leads to the expression
\begin{align}\label{eq:corr}
 & C_{n}({\bf x})= \\
 & \lim_{L\rightarrow\infty}\kern-.1em\Tr{e^{D(L-x_{n})}Me^{D(x_{n}-x_{n-1})}M\ldots Me^{D(x_{1}-0)}}.\nonumber
\end{align}
With $D$ we denote the transfer matrix $T$---introduced in Eq.~\eqref{eq:Liouvillian}---in its diagonal basis,
\be \label{eq:DX}
D=X^{-1}TX\,,
\ee
where the columns of $X$ represent the eigenvectors of $T$. Analogously, the matrix $M$ denotes $R_\beta^* \otimes R_\beta$ in the diagonal basis of $T$, 
\begin{equation}
	M=X^{-1}(R_\beta^* \otimes R_\beta)X. 
\end{equation}
Let us mention that the knowledge of $X$ is in principle not necessary to reconstruct the matrices $Q$ and $R$ and hence the according Lindblad equation~\cite{LongPaper}.

Specifically, the second- and third-order correlation functions take the form 
\be
\label{eq:C2}
C_2 (x) = \Tr{ \mathrm{e}^{D\infty}  M \textrm{e}^{D x} M}=\sum_{j=1}^{d^2}M_{1,j}M_{j,1}\textrm{e}^{\lambda_j x}
\ee
and 
\be
\label{eq:C3}
C_3 (x,x')  =\sum_{j,k=1}^{d^2}M_{1,k}M_{k,j}M_{j,1}\textrm{e}^{\lambda_j x}\textrm{e}^{\lambda_k (x'-x)},
\ee
with $\{\lambda_j\}$ being the eigenvalues of $T$. Due to the translation invariance of the system, we can set $x_1=0$. The tomographic possibilities of the cMPS formalism can be understood from Eqs.~(\ref{eq:C2}-\ref{eq:C3}): If the products $\{M_{1,k}M_{k,j}M_{j,1}\}$ are known, we can---using gauge arguments \cite{Wick}---require each of the matrix elements $\{M_{1,j}\}$ to be equal to one, which enables us to access each $M_{k,j}$ by dividing the appropriate terms:
\be
\frac{M_{1,k}M_{k,j}M_{j,1}}{M_{1,1}M_{1,j}M_{j,1}}= M_{k,j}.
\ee
Both numerator and denominator appear as coefficients in $C_3$ and can be determined with spectral estimation procedures. This means that in principle we just need to analyse a three-point function in order to obtain the building elements $M$ and $D$ of arbitrary-order correlation functions.

This reconstruction scheme demonstrates the central role of the matrices $M$ and $D$ to derive the different equivalent objects that describe the dynamics of an open quantum system: the Lindblad equation, the set of $n$-point correlation functions, the full counting statistics of the number of transferred particles and the cMPS state vector. These matrices $M$ and $D$ can therefore be considered as the central quantities on which our reconstruction machinery is based; this is illustrated in Fig.~\ref{fig:relations}.

\subsection{Use of the thermodynamic limit}
 
Intuitively, it is clear that the reconstruction of the matrices $M$ and $D$ should gain precision by increasing the number of correlation functions $C_n$ on which the reconstruction scheme is based. The same statement is valid when increasing the size of the set of available counting probabilities $P_n$. But in general, experiments will only provide us measurements of low-order correlation functions, typically those of the second- and third-order \cite{Gustavsson06, Gabelli13, Ubbelohde12}. A priori, this might render the reconstruction of the matrices $M$ and $D$ infeasible, but the work in Ref.~\cite{Wick} proved that this limitation can be circumvented by making use of the structure of the cMPS state vector combined with the thermodynamic limit. 

For a given finite region $I$ and a fixed bond dimension $d$, all expectation values can be computed from \textit{all} correlation functions $C_n({\bf x})$ taking values in the finite range $I$, ${\bf x}=(x_1,\dots, x_n)\subset I^{\times n}$. This contrasts with the situation  of having access to correlation functions $C_n({\bf x})$ for \textit{arbitrary} values of ${\bf x}=(x_1,\dots, x_n)$, but for low $n$. Here, arbitrary values ${\bf x}$ imply the thermodynamic limit, i.e. the finite region $I$ tends to infinity. Then indeed, low order correlation functions (typically $C_{1}, C_{2}, C_{3}$) are sufficient to reconstruct an arbitrary expectation value of an observable supported on $I$.

\section{cMPS reconstruction from low-order counting probabilities}
\label{sec:proof}

In this section, we extend the central role played by the matrices $M$ and $D$ for tomographic purposes by showing that they (more precisely: their equivalents $\mathcal{M}$ and $\mathcal{D}$) are also accessible from low-number detector-click statistics, i.e. the idle time probability density function $P_0$ and the density function $P_1$, which correspond to the detection of zero and one particle, respectively, within a certain time interval $\tau$. 

It is well-known that correlators and counting statistics are closely related. When assuming perfect detectors, the probability to observe $n$ events in the time interval between $t$ and $t+\tau$ is given \cite{KK64} by the expression
\begin{align}\label{eq:Pn0}
 & P_{n}(t,t+\tau)=\\
 &\nonumber \frac{1}{n!}\sum_{m=n}^{\infty}\frac{(-1)^{m-n}}{(m-n)!}\int_{t}^{t+\tau}\kern-1em\mathrm{d}t_{1}\cdots\int_{t}^{t+\tau}\kern-1em\mathrm{d}t_{m}\; C_{m}(t_{1},t_{2},\ldots,t_{m})\,,
\end{align}
where the correlation function $C_m$ has been introduced in Eq.~(\ref{eq:def_corr}). For a translationally invariant system, we can without loss of generality set $t=0$. Furthermore, when changing the integration bounds and performing the limit $L\rightarrow\infty$, we obtain
\begin{align}
\label{eq:Pn}
& P_n(\tau):=P_n(0,0+\tau)  = \nonumber\\
& \int_0^{\tau}\kern-.3em \mathrm{d}t_1 \int_0^{t_1}\kern-.3em \mathrm{d}t_2 \cdots \int_0^{t_{n-1}}\kern-1em \mathrm{d}t_n \; \tilde{C}_n(\tau,t_1,t_2,\ldots,t_n),
\end{align}
with $\tilde{C}_{n}(\tau,t_{1},\ldots,t_{n}):=$
\begin{align}
\label{eq:tildeCn}
&e_{1}^{T}Z^{-1}\mathrm{e}^{\mathcal{D}t_{n}}\mathcal{M}_{n}\cdots\mathcal{M}_{2}e^{\mathcal{D}(t_{1}-t_{2})}\mathcal{M}_{1}e^{\mathcal{D}(\tau-t_{1})}Ze_{1},
\end{align}
with the canonical unit vector $e_1$, the diagonal matrix $\mathcal{D}$ of $Q^* \otimes \one + \one \otimes Q$ with basis transformation matrix $Y$, $\mathcal{M}_j:=Y^{-1}(R_j^* \otimes R_j)Y$, and $Z:=Y^{-1}X$, where $X$ diagonalizes $T$ as defined in Eq.~\eqref{eq:DX}.
%
With Eqs.~(\ref{eq:Pn0}-\ref{eq:tildeCn}), the low-order counting probabilities $P_0(\tau)$ and $P_1(\tau)$ within a cMPS formulation are given by similar expressions to Eqs.~(\ref{eq:C2}) and (\ref{eq:C3}), namely 
\be
P_0(\tau) \!&=&\! e^T_1 Z^{-1} \mathrm{e}^{\mathcal{D}\tau}  Z e_1= \sum_{j=1}^{d^2}\hat{z}_{1,j}z_{j,1} \mathrm{e}^{\mu_j\tau}, \label{eq:P0}\\
P_1(\tau) \!&=&\! \sum_{j,k=1}^{d^2}\kern-.2em\hat{z}_{1,j}\mathcal{M}_{j,k}z_{k,1}\nonumber\\
&\times& \left((1-\delta_{j,k})\frac{\mathrm{e}^{\mu_k \tau} - \mathrm{e}^{\mu_j \tau}}{\mu_k-\mu_j}+\delta_{j,k}\tau \mathrm{e}^{\mu_j\tau}\right) \label{eq:P1},
\ee
with $\{\mu_j\}$ being the diagonal values of $\mathcal{D}$ and $\{z_{j,k}\}$ being the elements of the matrix $Z$ (with inverse $Z^{-1}=:(\hat{z}_{j,k})$). See Appendix~\ref{AppenA} for details.

As a first step, we can extract from $P_0$ and $P_1$ the coefficients $\{\hat{z}_{1,j}z_{j,1}\}$ and the eigenvalues $\{\mu_i\}$, which give rise to $\mathcal{D}$. The matrix elements of $\mathcal{M}$ can then in principle be determined using gauge arguments and under the assumption that the additive components of $P_n$ are linearly independent. From $\mathcal{M}$ and $\mathcal{D}$, the cMPS matrices $Q$, $R$, and $K$ describing the dynamics of the open quantum system can be determined in a straightforward way (see Appendix~\ref{AppenA} for details).


Let us comment on the feasibility of this reconstruction scheme with present technology. In order to measure $P_0$ and $P_1$, efficient single-particle detectors without \emph{dark-counting} and \emph{tiny dead-time} are necessary. \emph{Dark-counting} leads to detector output pulses in the absence of any incident photons while the \emph{dead-time} is the time interval after a detection event during which the detector can not detect another particle. Although significant experimental efforts have been made in order to improve single-photon \cite{Eisaman11} and single-electron detectors \cite{Thalineau14, Gasparinetti15}, the state-of-the-art for single-particle detection is not yet sufficient to perform a reliable measurement of $P_1$. For the moment, these experimental constraints make the reconstruction scheme based on $P_1$ only valid on a formal, mathematical level. In the light of the recent experimental progress towards the reliable detection of single particles, we believe that this idea will become relevant in the future.



\section{Application to fermionic quantum transport experiments}
\label{sec:QT}

Very recent works have successfully formulated experimental setups in cavity QED and ultra-cold Bose gases as well as the corresponding measurements in terms of cMPS \cite{cMPSTomographyShort, CavityQED}. This allowed them to make predictions for higher-order correlation functions that are not accessible experimentally and to investigate the ground-state entanglement. 

Here, we tackle the problem of formulating quantum transport experiments and the corresponding measurements (average charge current, charge noise) in cMPS terms. To this end, we demonstrate that the field that is leaking out and is measured in a quantum transport experiment belongs to the cMPS variational class. We then provide an example to illustrate the equivalence between an Hamiltonian and a cMPS formulation by considering one of the simplest transport  experiment, namely single electrons tunnelling through a single-level quantum dot. We derive the first-order and second-order correlation functions in cMPS terms, and show that we recover the well-known expression of the average current and charge noise, when writing the cMPS state Eq.~(\ref{eq:def_cmps}) in terms of the parameters of the quantum system.

\subsection{Quantum transport experiments \\ in terms of cMPS}

We now turn to a description of the physical setting under consideration. We assume here transport experiments, where single electrons transit through a scatterer coupled to fermionic reservoirs. The reservoirs, considered at equilibrium, are characterised by their chemical potential and their temperature via the Fermi distribution. The bias energy and the bias temperature between the different reservoirs will set the direction of the charge current. For the sake of simplicity, we restrict ourselves to two reservoirs, the source and the drain. \ger{This transport setting can be described by a tight-binding Hamiltonian,}
\be
\label{eq:Ht}
\hat{H}_\mathrm{T} = \hat{H}_\text{sys} + \hat{H}_\text{res}  + \hat{H}_\text{int},
\ee
where $ \hat{H}_\text{sys}$ relates to the quantum system under investigation, which acts as scatterer. It is characterized by discrete energy levels $\varepsilon_i$ with occupation number operators given by $\hat{d}^\dagger_{i , \sigma} \hat{d}_{i , \sigma}$ ($\hat{d}_{i , \sigma}$ and $\hat{d}_{i , \sigma}^\dagger$ denote the fermionic annihilation and creation operators for an electron on the energy level $i$ and spin degree of freedom $\sigma =\,\, \uparrow,\, \downarrow$). 
The Hamiltonian $\hat{H}_\text{res}$ relates to the left and right reservoirs, and $\hat{H}_\text{int}$ describes the interaction between the quantum system and the reservoirs,
\begin{align}
&\hat{H}_\text{res} =  \sum_{\alpha=L,R} \sum_{\sigma = \uparrow, \downarrow} \int_{0}^{E_{\alpha}} \mathrm{d}E \, E \, \hat{c}^\dagger_{\alpha, \sigma}(E) \, \hat{c}_{\alpha, \sigma}(E)\,, \label{eq:Hleads} \\
&\hat{H}_\text{int} = \sum_{\alpha=L,R}  \sum_{i, \sigma} \int \mathrm{d}E  \left( t_{\alpha,  i,  \sigma}(E)\,   \hat{d}_{i , \sigma} \otimes \hat{c}^\dagger_{\alpha , \sigma}(E) + \mathrm{h.c.} \right)\,. \label{eq:Hint}
\end{align}
The creation and annihilation operators of the reservoirs, $\hat{c}_{\alpha}$ and $\hat{c}^\dagger_{\alpha}$, satisfy the canonical anti-commutation relations and $\alpha = L,R$ denotes the left and right reservoirs respectively. The amplitude $t_{\alpha,  i,  \sigma}$ sets the interaction between the quantum system and its environments.

In order to model a DC source, the energy levels in the left and right reservoirs are assumed to be densely filled up to the energies $E_\mathrm{F} + eV$ and $E_\mathrm{F}$, respectively. Here, $E_\mathrm{F}$ is the Fermi energy and $V$ is the bias potential applied on the "source" reservoir. At zero temperature, the bias energy $eV$ enables uni-directional transport of electrons between the left and right reservoirs. It plays a similar role to the frequency bandwidth when, e.g., considering cavity QED setups, and fixes the energy domain over which electronic transport takes place.

With this assumption about the direction of propagation of the electrons (from left to right), we will see that Eq.~(\ref{eq:Ht}) is equivalent to a generalised version of the cMPS Hamiltonian introduced in Refs.~\cite{cMPS1,cMPS2},
\be
\label{eq:Hcmps}
\hat{H}_\text{cMPS} =  Q \otimes \hat{\one} +  \left( R_{\mathrm{L}} \otimes \hat{\psi}_{\mathrm{L}}  +R_{\mathrm{R}} \otimes \hat{\psi}_{\mathrm{R}}^{\dagger} \right)\,,
\ee
where the matrices $Q$ and $\{R_\alpha\}$ and the quantum fields $\{\hat{\psi}_\alpha\}$ have been introduced in Section~\ref{sec:cmps_tomo}. The cMPS Hamiltonian for quantum transport experiment reflects the direction of the current: a fermionic excitation present on the left of the scatterer is annihilated at the scatterer as described by the quantum field $\hat{\psi}_\mathrm{L}$ (an electron jumps into the scatterer). Similarly, a fermionic excitation present on the right of the scatterer is created at the scatterer as described by the quantum field $\hat{\psi}^\dagger_\mathrm{R}$ (an electron jumps out of the scatterer). The case of a multi-terminal setup can be considered in a similar way. Showing that Eqs.~\eqref{eq:Ht} and \eqref{eq:Hcmps} are equivalent implicates that there is a fermionic quantum field leaking out of the scatterer to be measured and that it belongs to the cMPS variational class. Such a description of the transport experiment corresponds to a fermionic version of the input-output formalism of cavity-QED setups.

\begin{figure*}
\centering
\includegraphics[width=0.75\textwidth]{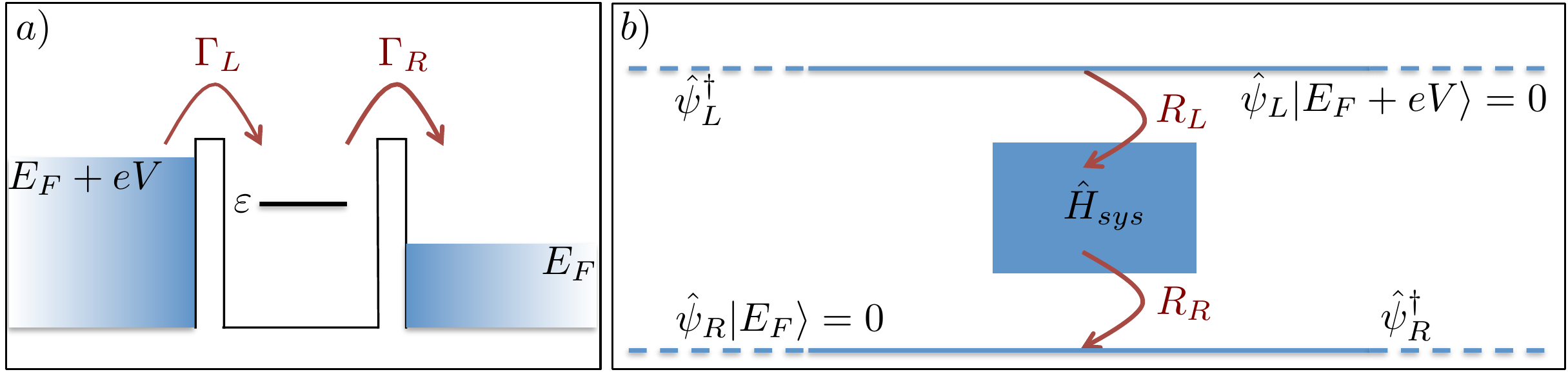}
\caption{(Color online). Scheme of a transport experiment through a single-level quantum dot. a) The single-level quantum dot with energy $\varepsilon$ is tunnel-coupled to two biased reservoirs with coupling strengths $\Gamma_\mathrm{L}$ and $\Gamma_\mathrm{R}$. Spin-less single-electron tunnelling events take place in the energy window $eV$ above the Fermi sea with energy $E_\mathrm{F}$. b) The same transport experiment from the open quantum system perspective for a cMPS formulation. The single-level dot is described by $\hat{H}_\text{sys}$ and coupled to fermionic quantum fields $\hat{\psi}_\mathrm{L}$ and $\hat{\psi}_\mathrm{R}$. The coupling matrices $R_\mathrm{L/R}$ depend on the parameters $\Gamma_\mathrm{L/R}$, see Eqs.~(\ref{eq:Heff0}-\ref{eq:Heff})). The transport direction fixed by the biased energy between the left and right reservoirs is ensured in the cMPS formulation by imposing $\hat{\psi}_\mathrm{L} \vert E_\mathrm{F} + eV \rangle = 0$ and $\hat{\psi}_\mathrm{R} \vert E_\mathrm{F} \rangle = 0$.}
\label{fig:setup}
\end{figure*}

Using Eq.~(\ref{eq:Hint}), the quantum field leaking out of the quantum system, $\hat{\psi}_{\mathrm{R}}(t)$, can be written in terms of the creation operator in the right reservoir $\hat{c}_{\mathrm{R}}$\ad{;} the incoming quantum field can be written in a similar way in terms of the creation operator in the left 
reservoir $\hat{c}_{\mathrm{L}}$,
\be
\label{eq:psi}
\hat{\psi}^\dagger_{\alpha , \sigma}(t) = \int_E \mathrm{d}E \; e^{-i E t/\hbar} \hat{c}^\dagger_{\alpha , \sigma}(E), \quad \alpha=\mathrm{L,R}\,.
\ee
The Fermi sea for the electrons is taken into account in the following way: On the right side of the scatterer, the quantum field satisfies $\hat{\psi}_R(t)\ket{E_\mathrm{F}}= 0$, where $ \ket{E_\mathrm{F}}$ denotes the state of the Fermi sea at energy $E_\mathrm{F}$, whereas on the left side of the scatterer, $\hat{\psi}_\mathrm{L}(t)  \ket{E_\mathrm{F} + eV} =0$, where the state vector $\ket{E_\mathrm{F} + eV}$ defines the state of a Fermi sea at energy $E_\mathrm{F} + eV$.

Assuming that the energy levels $\varepsilon_i$ of the quantum system are well inside the bias energy window $eV$, we can rewrite the integration over the energy domain $E$ as $\int_E \mathrm{d}E = \int_0^{eV} \mathrm{d}E$.

This assumption is the so-called \emph{large-bias limit}, which is considered in order to derive the master equation corresponding to the tight-binding Hamiltonian. In quantum optics, it corresponds to a finite frequency bandwidth, which allows the use of the rotating wave approximation \cite{Gardiner, CavityQED}. In the following, we assume that the interaction amplitude is spin- and energy-independent within the interval $[E_\mathrm{F},E_\mathrm{F}+eV]$ :  $t_{\alpha,  i,  \sigma}(E) = t_\alpha$. Let us remark that the demonstration remains valid with an interaction amplitude that depends on spin and energy.

In a rotating frame with respect to the energies of the reservoirs and after a Jordan-Wigner transformation using the definitions of the quantum fields $\hat{\psi}_{\mathrm{R,L}}$ given in Eq.~\eqref{eq:psi}, the Hamiltonian in Eq.~\eqref{eq:Ht} can be rewritten as
\be
\hat{H}_\mathrm{T} = \hat{H}_\text{sys} \otimes \hat{\one} + \sum_{\alpha=\mathrm{L,R}} \sum_{i, \sigma} 
\left( t_{\alpha} \hat{d}_i \otimes \psi^\dagger_{\alpha , \sigma}(t)+ \mathrm{h.c.}\right).
\ee
Following quantum optics calculations---which remain valid in this case because $\hat{H}_T$ is a transport version of the spin-boson model---we finally arrive at an effective non-Hermitian Hamiltonian
\be
&&\hat{H}_\text{eff} = \left(\hat{H}_\text{sys} - \frac{i \hbar}{2} \sum_{\alpha=\mathrm{L,R}} \sum_{i, \sigma} \Gamma_{\alpha} \hat{d}_{i , \sigma}^\dagger \hat{d}_{i , \sigma} \right) \otimes \one \label{eq:Heff0} \\
&&+ \sum_{i, \sigma} \left( \sqrt{\Gamma_\mathrm{R}} \hat{d}_{i , \sigma} \otimes \hat{\psi}_{\mathrm{R} , \sigma}^\dagger(t) + \sqrt{\Gamma_\mathrm{L}} \hat{d}_{i , \sigma}^\dagger \otimes \hat{\psi}_{\mathrm{L} , \sigma}(t) \right) \nonumber 
\ee
with $t_{\alpha} := \sqrt{\Gamma_\alpha}$. Expressed in the eigenbasis of $\hat{H}_\text{sys}$, the operators $\sqrt{\Gamma_\mathrm{R}} \hat{d}_{i , \sigma}$ and $\sqrt{\Gamma_\mathrm{L}} \hat{d}^\dagger_{i , \sigma}$ take the form of matrices labelled $R_{\mathrm{R}, i, \sigma}$ and $R_{\mathrm{L} ,i, \sigma}$, respectively. The effective non-Hermitian Hamiltonian can then be rewritten in a compact form
\begin{eqnarray}
\label{eq:Heff}
\hat{H}_\text{eff}  &=& Q \otimes \hat{\one} \\
&+& \sum_{i, \sigma} \left( R_{\mathrm{L} ,i, \sigma}^\dagger \otimes \hat{\psi}_{\mathrm{L} , \sigma}(t) + R_{\mathrm{R}, i, \sigma} \otimes \hat{\psi}_{\mathrm{R} , \sigma}^\dagger(t)  \right)\,.\nonumber
\end{eqnarray}
When comparing this effective Hamiltonian with Eq.~\eqref{eq:Hcmps}, the identification of the matrix $Q$ and the matrices $\{R_{\alpha}\}$ is direct. For spin-less fermions, the matrices $R$ verify $R_{\alpha, i,\sigma}^2 =0$ in order to satisfy the Pauli principle.
\ger{Eq.\ \eqref{eq:Heff} demonstrates that transport settings can be adequately formulated within the cMPS framework. This result is important as it clears the way for applying methods from cMPS tomography to fermionic quantum transport experiments.}

\subsection{Single energy-level quantum dot}\label{sub:QD}

To illustrate the input-output formalism and the cMPS formulation of quantum transport experiments, we consider one of the simplest setups, namely a single energy-level quantum dot, without spin-degree of freedom, weakly coupled to two fermionic reservoirs. Even though this experiment is \je{characterised by Markovian dynamics}, this example is of particular interest for this work as it has been widely investigated experimentally. In Section~\ref{sec:WTD}, we will use real data obtained in Ref.~\cite{Ubbelohde12} for this setup to show that cMPS tomography allows us to access the electronic distribution of waiting times. 

This simple transport experiment is sketched in Fig.~\ref{fig:setup} and the corresponding Hamiltonian reads
\be
\hat{H}_T &=& \varepsilon \,  \hat{d}^\dagger \hat{d} +  \sum_{\alpha=L,R} \int \mathrm{d}E  \left( \sqrt{\Gamma_\alpha}\,   \hat{d}\otimes  \hat{c}^\dagger_{\alpha}(E) + \mathrm{h.c.} \right) + \hat{H}_\text{res}\,. \nonumber \\
&&
\ee
Because of the weak coupling assumption between the dot and the reservoirs, the tight-binding Hamiltonian $\hat{H}_T$ reduces here to a tunnelling Hamiltonian.
Assuming that we perform a measurement on the right of the scatterer, the first two correlation functions of the right quantum field $\hat{\psi}_\mathrm{R}(t)$ read in terms of cMPS matrices
\be
\label{eq:1corr}
\langle \hat{\psi}_\mathrm{R}^\dagger \hat{\psi}_\mathrm{R} \rangle &=& \lim_{L\rightarrow\infty} \text{Tr} \left[ e^{TL} (R^*_\mathrm{R}\otimes R_\mathrm{R})\right]
\ee
and
\begin{multline}
\label{eq:2corr}
\langle \hat{\psi}_\mathrm{R}^\dagger(0) \hat{\psi}_\mathrm{R}^\dagger(\tau) \hat{\psi}_\mathrm{R}(\tau)\hat{\psi}_\mathrm{R}(0)\rangle \\
= \lim_{L\rightarrow\infty} \text{Tr} \left[ e^{T(L-
\tau)} (R^*_\mathrm{R}\otimes R_\mathrm{R}) e^{T \tau} (R^*_\mathrm{R}\otimes R_\mathrm{R}) \right].
\end{multline}
The matrices $R_{\mathrm{R/L}}$ correspond to the operators $\sqrt{\Gamma_\mathrm{R}} \hat{d}$ and $\sqrt{\Gamma_\mathrm{L}} \hat{d}^\dagger$ expressed in the eigenbasis of the single-level quantum dot, $\{ \vert 0 \rangle, \vert 1 \rangle\}$ (empty and occupied state),
\be
R_\mathrm{L} = \left( \begin{array}{cc} 
0 & 0 \\
\sqrt{\Gamma_\mathrm{L}} & 0
\end{array} \right)\,, \quad R_\mathrm{R} = \left( \begin{array}{cc} 
0 & \sqrt{\Gamma_\mathrm{R}} \\
0 & 0
\end{array} \right)\,.
\ee

Inserting these expressions into Eq.~\eqref{eq:1corr}, we recover the well-known expression for the steady-state current of a single-level QD coupled to biased reservoirs \cite{Stoof95, Blanter00},
\be
\label{eq:current}
\langle \hat{\psi}_\mathrm{R}^\dagger \hat{\psi}_\mathrm{R}\rangle &=&\frac{\Gamma_\mathrm{L} \Gamma_\mathrm{R}}{\Gamma_\mathrm{R} + \Gamma_\mathrm{L}} =:\langle \hat{I} \rangle_\mathrm{ss}.
\ee
Furthermore, we can derive the noise spectrum from Eq.~(\ref{eq:2corr}) via the MacDonald formula \cite{MacDonald49, Flindt05, Lambert07}
\be
\label{eq:noise}
S(\omega) = 2 \langle \hat{I} \rangle_\mathrm{ss} \left( 1-\frac{2 \Gamma_\mathrm{L} \Gamma_\mathrm{R}}{(\Gamma_\mathrm{L}+\Gamma_\mathrm{R})^2 + \omega^2} 
\right).
\ee
This example aims at bridging the gap between a more traditional Hamiltonian and the cMPS formulation, which allows to write these well-known expressions in terms of the parameter matrices $Q$, $T$, and $\{R_\alpha\}$.




\section{Reconstruction of waiting time statistics}
\label{sec:WTD}

In this section, we address the problem of accessing the distribution of waiting times in electronic transport experiments. As mentioned in the introduction, a direct measurement of the WTD is not yet possible due to the lack of single-particle detectors with sufficient accuracy. Here, we propose to reconstruct the WTD based on the experimental measurements of low-order correlation functions. The reconstruction is carried out using the machinery based on cMPS presented in Section~\ref{sec:cmps_tomo} and the formulation of transport experiments in terms of cMPS as exposed in Section~\ref{sec:QT}.


\subsection{Definitions}

The statistics of waiting times can be expressed in terms of the probability density function $P_0$, which---as a function of $\tau$---expresses the probability of having detected zero particles in the interval $[0,\tau]$. In terms of $P_0$, the WTD has first been derived in the context of quantum transport experiments in Ref.~\cite{Albert12},
\be\label{eq:W}
\mathcal{W}(\tau) = \langle \tau \rangle \frac{\partial^2 P_0(\tau)}{\partial \tau^2}\,.
\ee
Here, $\langle \tau \rangle$ denotes the mean waiting time. Inserting $P_0(\tau)$ in cMPS terms (Eq.~\eqref{eq:P0}), we arrive at an expression for $\mathcal{W} $ in terms of the cMPS matrices $D$, $\mathcal{D}$ and $Z$ defined in Eqs.~(\ref{eq:DX}) and (\ref{eq:tildeCn}),
\be
\label{eq:WTD}
\mathcal{W}(\tau) = \frac{1}{c}e_1^T(D^2Z^{-1}-2DZ^{-1}\mathcal{D}+Z^{-1}\mathcal{D}^2)\textrm{e}^{\mathcal{D}\tau}
Ze_1.\nonumber\\
\ee
The normalisation factor $c>0$ ensures that $\int_0^\infty \mathcal W(\tau) \mathrm{d}\tau=1$. 
Equation~(\ref{eq:WTD}) allows us to access the WTD from the measurements of the low-order correlation functions via the use of the cMPS machinery to reconstruct the cMPS matrices $D$, $\mathcal{D}$ and $Z$.


\subsection{Results based on experimental data}

\begin{figure}
\centering
\includegraphics[width=.9\columnwidth]{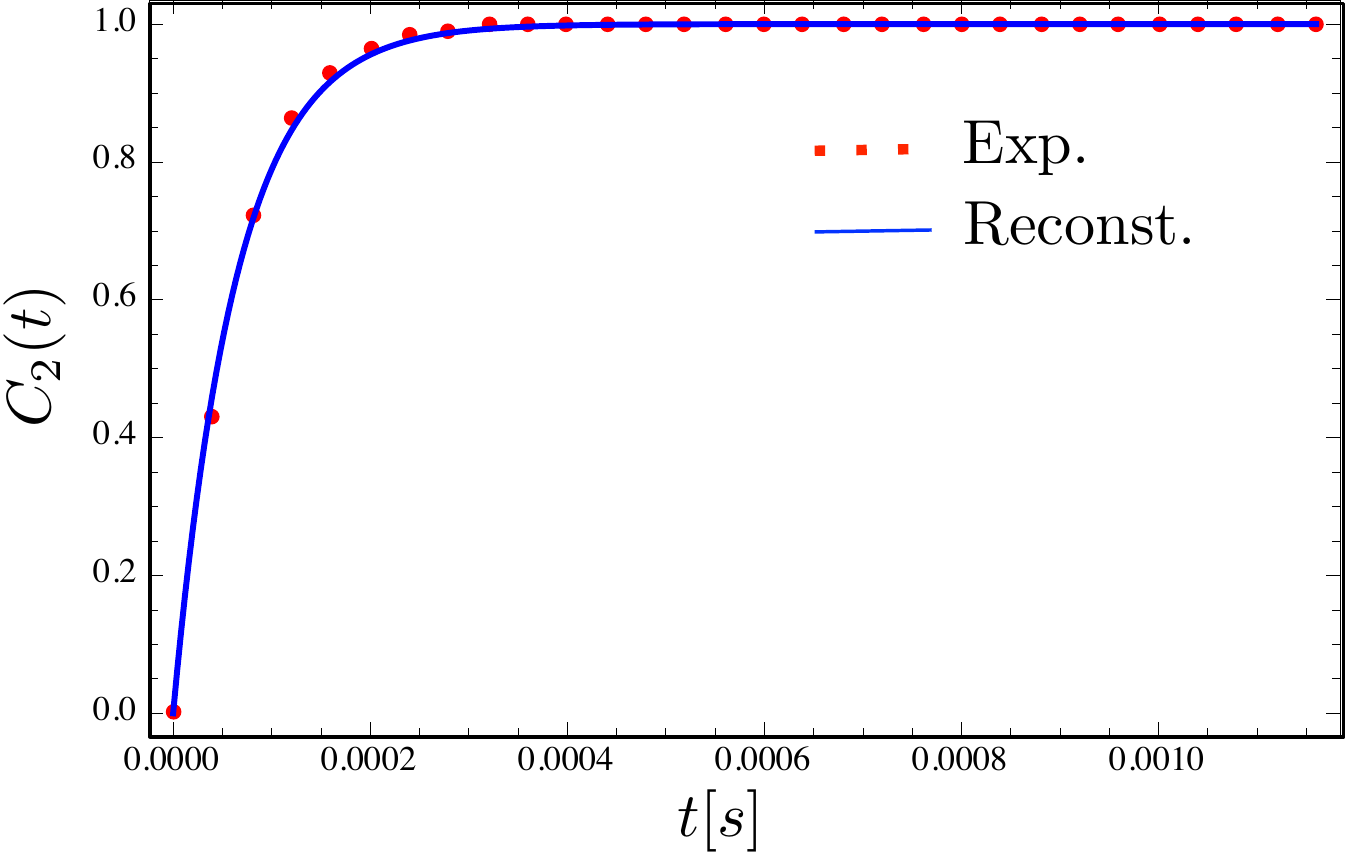}
\caption{(Color online). Two-point correlation function $C_2$ for single fermions tunnelling through a single-level quantum dot with data from Ref.~\cite{Ubbelohde12} (red-dotted curve). The blue curve, obtained from the reconstructed values of the parameters $\Gamma_\mathrm{L,R}$ using a cMPS formulation of the quantum experiment, agrees well with the experimental measurement of $C_2$. The deviation for small times is due to experimental limitations in the time bin with respect to which the current trace is resolved.}
\label{fig:C2}
\end{figure}


We demonstrate our novel approach to derive the WTD from the measurement of correlation functions using experimental data obtained in Ref.~\cite{Ubbelohde12} for spinless electrons tunnelling through a single-level quantum dot. This system is also known as a single-electron transistor at the nanoscale and has been discussed in Section~\ref{sub:QD}. The experiment in Ref.~\cite{Ubbelohde12} has been carried out in the kHZ frequency range, where a time-resolved measurement of the current trace is possible. Although all the statistics---including correlation functions of arbitrary order as well as the WTD---can directly be computed from this time-resolved current trace, this experiment provides an ideal test-bed for our proposal. We can compare the WTD obtained from our reconstruction scheme based on cMPS with the WTD directly deduced from the experimental current trace. 

Due to the simplicity of the setup, our proposed method to access the WTD only requires the two-point function $C_2$. This one can directly be derived from the experimental spike train $I$ (the time-resolved current trace) and is shown in Fig.~\ref{fig:C2} (red dots). The rates $\Gamma_\mathrm{L} = 13.23$ kHz and $\Gamma_\mathrm{R} = 4.81$ kHz have been determined experimentally and the corresponding $C_2$-function agrees very well with the analytical expression when the detector rate is taken into account\cite{Ubbelohde12}
\be
\label{eq:C2_SET}
C_2(\tau)=\frac{\Gamma_\mathrm{L}\Gamma_\mathrm{R}}{\Gamma_\mathrm{L}+\Gamma_\mathrm{R}}\left(1-\mathrm{e}^{-\tau(\Gamma_\mathrm{L}+\Gamma_\mathrm{R})}\right)\,.
\ee

In our reconstruction scheme, the quantity $\Gamma_\mathrm{L}+\Gamma_\mathrm{R}$ can be determined from the current spike train autocorrelation function $I\star I$ by least squares methods or spectral estimation procedures analogous to the procedure described in Ref.~\cite{LongPaper}. By requiring $\Gamma_\mathrm{L}>\Gamma_\mathrm{R}$ and using the expression of the steady-state current (see Eq.~(\ref{eq:current})), $\Gamma_\mathrm{L}$ and $\Gamma_\mathrm{R}$ can be uniquely identified. The reconstructed values for the rates are 
\be
\Gamma_\mathrm{L}^\text{recon} &=& 10.80 \text{kHz} \label{eq:GL} \,,\\
\Gamma_\mathrm{R}^\text{recon} &=& 4.76\, \text{kHz} \label{eq:GR}\,.
\ee
The differences to the values from Ref.~\cite{Ubbelohde12} are well within the range we would expect, regarding the time-resolution in the spike train data. The curve plotted from these reconstructed values of the parameters $\Gamma_\mathrm{L,R}$ is shown in Fig.~\ref{fig:C2} in blue. The slight deviation between the experimental points and this reconstructed $C_2$-function is due to the discretisation of the counting time intervals used in the experiment: The size of each time bin is not much smaller than the time scale on which $C_2$ changes mostly. This leads to an error in the estimation of the damping factor $\Gamma_\mathrm{L}+\Gamma_\mathrm{R}$ and explains the difference of the blue and the red dotted curves. Naturally, one could expect a more accurate reconstruction of the parameters $\Gamma_\mathrm{L}$ and $\Gamma_\mathrm{R}$ when increasing the time resolution of the current trace or of the measurement of $C_2$.

From $\Gamma_\mathrm{L}$ and $\Gamma_\mathrm{R}$, the corresponding cMPS matrices $R_\mathrm{L}$ and $R_\mathrm{R}$ can be constructed, as well as the matrices $M$ and $D$. In this simple case, we did not need to employ the whole reconstruction procedure from Ref.\ \cite{LongPaper}. Indeed, it is clear from Eq.~(\ref{eq:C2_SET}) that only two out of the four parameters that characterise the system appear: $C_n$ only depends on the tunnelling rates $\Gamma_\mathrm{L}$ and $\Gamma_\mathrm{R}$---the eigenenergies $0$ and $\varepsilon$ of $\hat{H}_\text{sys}$ do not contribute \footnote{The eigenvectors of $T$ only depend on $\Gamma_\mathrm{L}$ and $\Gamma_\mathrm{R}$, this applies to $M_\mathrm{L}$, $M_\mathrm{R}$ and all residues as well. Accordingly, the two non-real poles are the only quantities that depend on $\epsilon$, however, only the residues connected to the two real poles do not vanish. When adding off-diagonal elements to $K$, the terms mix and a dependency on $K$ arises.}. This will in general not be the case.

\begin{figure}
\centering
\includegraphics[width=.9\columnwidth]{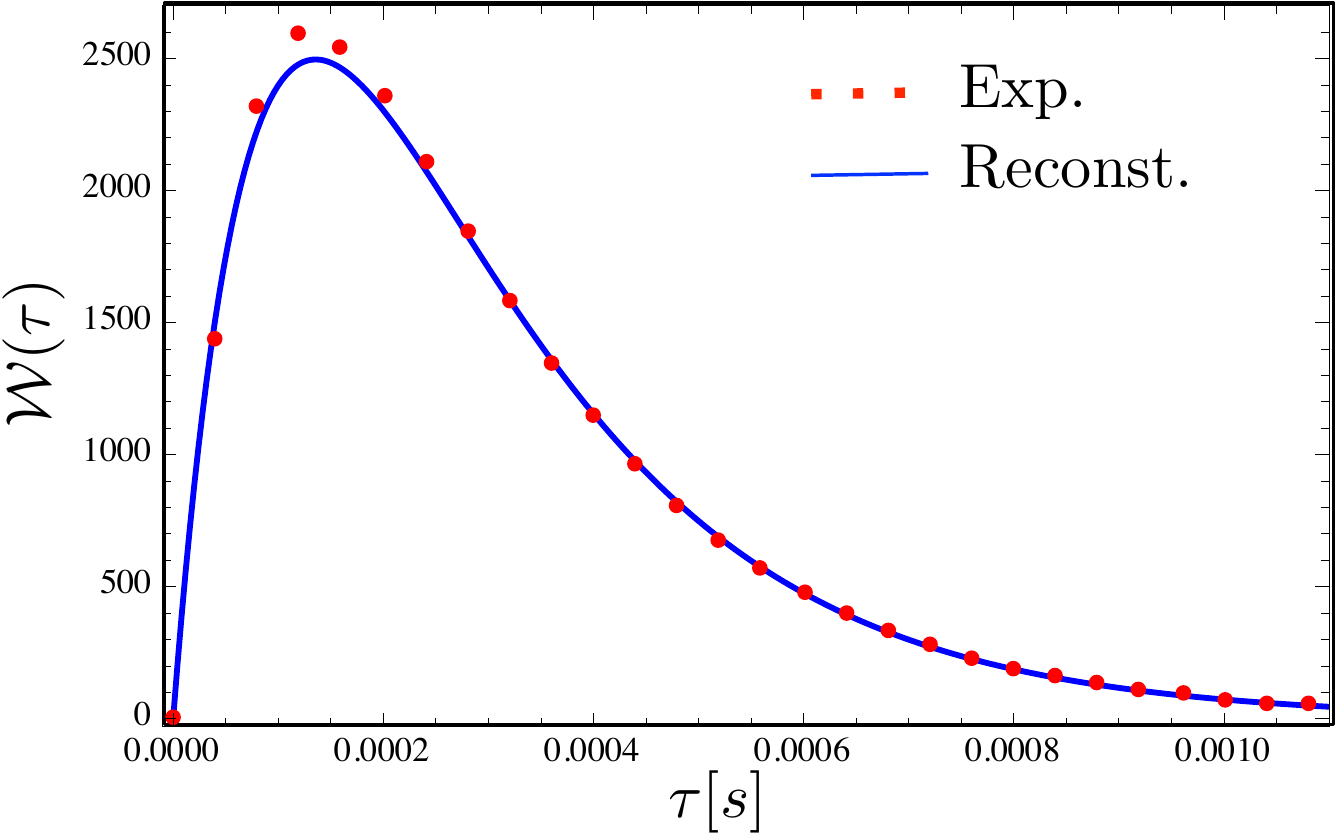}
\caption{(Color online). WTD obtained from state-of-the-art experimental measurements with data from Ref.~\cite{Ubbelohde12}. The reconstructed WTD using Eq.~(\ref{eq:WTD}) is shown in blue. It matches the WTD obtained directly from the time resolved experimental current trace (red dots) well. The deviation is due to the finite-sized time bin corresponding to the resolution of the current trace. A more accurate reconstruction of the WTD is expected by  increasing the time resolution of the current trace or of the measurement of $C_2$.
}
\label{fig:wtd}
\end{figure}

The matrices $R_\mathrm{L}$ and $R_\mathrm{R}$ give access to the matrices $D$ and $\mathcal{D}$ by direct computation. Inserting the latter into Eq.~(\ref{eq:WTD}), the WTD can be reconstructed and the result is plotted in Fig.~\ref{fig:wtd} (blue curve). In order to build confidence in our procedure, we compare this result with the experimentally-accessible WTD (red dots).
Let us recall that the transport rate is in the kHz range, hence the WTD can directly be extracted from the current spike train $I$: By sorting, counting all (discrete) waiting times between two consecutive incidents, and subsequently normalising the resulting histogram, one obtains the red-dotted WTD in Fig.~\ref{fig:wtd}. The slight deviation between the WTD reconstructed via our proposal and the experimental one is again due to the discretisation of the counting time intervals. One could expect a more accurate reconstruction of the WTD when increasing the time resolution of the current trace.

Although the WTD shown in Fig.~\ref{fig:wtd} reflects the most elementary transport properties of single independent particles through the Poissonian distribution, it bridges the gap between theoretical predictions and experiments. It demonstrates that our reconstruction procedure based on a cMPS formulation of an open quantum fermionic system is reliable to access the WTD from the measurements of low-order correlation functions. This opens the route to access the WTDs in the quantum regime from low-order correlation-functions measurements in the high-frequency domain.

\section{Conclusion}

In this work, we have taken an approach motivated by continuous matrix product states to perform tomographic reconstructions of quantum transport experiments. On a formal level, we have extended this formalism to perform a reconstruction of unknown dissipative processes based on the knowledge of low-order counting probabilities. We then demonstrated that continuous matrix product states is an adequate formalism to describe quantum transport experiments based on tight-binding Hamiltonians.

This work advocates a paradigm change in the analysis of transport experiments. The traditional method is to make explicit use of a model to put the estimated quantities into context, a model that may or may not precisely reflect the physical situation at hand. The cMPS approach is to not assume the form of the model, with the exception that the quantum state can be described by a cMPS. Such an approach is of particular interest as it opens the way to the access of quantities that are not measurable experimentally with current technologies, high-order correlation functions and distributions of waiting times.

To convincingly demonstrate the functioning of cMPS tomographic tools applied to quantum transport experiments, we presented a simple example that consists of electrons tunnelling through a single-level quantum dot. Making use of experimental data, we showed that we could successfully reconstruct the distribution of waiting times from the measurement of the two-point correlation function only. This work constitutes therefore a significant step towards accessing the waiting time distribution in the quantum regime experimentally, a challenge present for several years now. Importantly, the application of our reconstruction procedure goes beyond the interest in waiting time distributions: It also provides an access to higher-order correlation functions, which are key quantities to better understand interacting quantum systems.

In subsequent research, it would be desirable to further flesh out the statistical aspects of the problem. After all, the description in terms of continuous matrix product states constitutes a statistical model. It would constitute an exciting enterprise in its own right to identify region estimators that provide efficiently computable and reliable confidence regions \cite{Statistics} when considering the problem as a statistical estimation problem, related to the framework put forth in Refs.\ \cite{Robin,Christandl,Errors}. We hope that the present work inspires such further studies of transport problems in the mindset of quantum tomography.

\section*{Acknowledgements}

We acknowledge the group of R. Haug, and especially N. Ubbelohde, for sharing with us the experimental data. We also acknowledge valuable comments and discussions with M. Albert, D. Dasenbrook and C.\ Flindt. G.\ H. acknowledges support from the A.\ von Humboldt foundation, from the Swiss NCCR QSIT and from the ERC grant MesoQMC, J.\ E. from the BMBF (Q.com), the EU (RAQUEL, SIQS, AQuS, COST), and the ERC 
grant TAQ.


\appendix

\section{Reconstruction method from low-order counting probabilities}
\label{AppenA}

In this appendix, we provide further technical details on the reconstruction scheme based on the measurement of low-order counting probabilities, $P_0$ and $P_1$. 
 The goal is to access the central cMPS parameter matrices $\mathcal{M}$ and $\mathcal{D}$. 
We refer to Fig.~\ref{fig:relations} for a general view of the reconstructible items.

We start from Eq.~\eqref{eq:Pn0} in the main text. By changing the integration bounds, we obtain Eq.~\eqref{eq:Pn},
\begin{align*}
 P_n(\tau)= 
 \int_0^{\tau}\kern-.3em \mathrm{d}t_1 \int_0^{t_1}\kern-.3em \mathrm{d}t_2 \cdots \int_0^{t_{n-1}}\kern-1em \mathrm{d}t_n \; \tilde{C}_n(\tau,t_1,t_2,\ldots,t_n),
\end{align*}
where the integrand $C_n$ is altered to 
\begin{align}\label{eq:A1}
 & \tilde{C}_{n}(\tau,t_{1},\ldots,t_{n}):=\lim_{L\rightarrow\infty}\mathrm{Tr}\big[\mathrm{e}^{T(L-t_n-\tau)}\mathrm{e}^{St_{n}}R_{n}^{*}\otimes R_{n}\cdots\nonumber \\
 & \qquad\qquad\,\,\cdots R_{2}^{*}\otimes R_{2}\mathrm{e}^{S(t_{1}-t_{2})}R_{1}^{*}\otimes R_{1}\mathrm{e}^{S(\tau-t_{1})}\big].
\end{align}
Note that in contrast to Eq.~\eqref{eq:corr} where the propagating matrix is the transfer matrix $T$ defined by Eq.~(\ref{eq:Liouvillian}) (or equivalently its diagonal representation $D$), the propagating matrix in the exponential terms between two measurement points now is the matrix $S$, which is defined by 
\be
S:=Q^* \otimes \one + \one \otimes Q=T-\sum_j R_{j}^{*}\otimes R_{j}\,.
\ee
We can further simplify Eq.~\eqref{eq:A1} by performing the thermodynamic limit $L\rightarrow\infty$. The spectrum of $T$ for a generic system consists of complex values with negative real part and only one eigenvalue being equal to zero. When taking the limit $L\rightarrow\infty$, all eigenvalue contributions to $\mathrm{e}^{T(L-\tau)}$ vanish, except the one corresponding to the zero eigenvalue. Hence
\be
\lim_{L\rightarrow \infty}\mathrm{e}^{T(L-\tau)}=Xe_1 e^T_1X^{-1}\,,
\ee
with the first canonical unit vector denoted by $e_1$ and the basis transformation matrix $X$ to the diagonal basis of $T$. Similar to $D$ and $T$, we define the matrix $\mathcal{D}$ as the diagonal matrix of $S$ with basis transformation matrix $Y$
\be
S=Y \mathcal{D} Y^{-1}\,.
\ee
By defining the matrices $\mathcal{M}_j:=Y^{-1}(R_j^* \otimes R_j)Y$ for $ j=1,\dots,n$, and setting $Z:=(z_{j,k}):= Y^{-1} X$ (with inverse $Z^{-1}=:(\hat{z}_{j,k})$), we arrive at Eq.~\eqref{eq:tildeCn},
\begin{align*}
&\tilde{C}_{n}(\tau,t_{1},\ldots,t_{n})=\\
& e_{1}^{T}Z^{-1}\mathrm{e}^{\mathcal{D}t_{n}}\mathcal{M}_{n}\cdots\mathcal{M}_{2}\mathrm{e}^{\mathcal{D}(t_{1}-t_{2})}\mathcal{M}_{1}\mathrm{e}^{\mathcal{D}(\tau-t_{1})}Ze_{1}.
\end{align*}
Eq.~\eqref{eq:tildeCn} and Eq.~\eqref{eq:corr} have a close structural resemblance: the matrices $M$ and $\mathcal{M}$ are \emph{similar} in the linear algebra sense, i.e., there exists a basis transformation from $\mathcal{M}$ to $M$. The matrices $D$ and $\mathcal{D}$  are the diagonal matrices of the transfer matrix $T$ and the matrix $S$, respectively. It is straightforward to transform $M$ and $D$ into $\mathcal{M}$ and $\mathcal{D}$ and vice versa: By subtracting $M$ from $D$, we obtain $S$ (up to similarity/basis transformation), whose diagonal matrix is $\mathcal{D}$. Applying the same basis transformation (from $(D-M)$ to $\mathcal{D}$) to the matrix $M$ results in the matrix $\mathcal{M}$.

For $n=0$, the counting probability function then reads
\begin{equation}
P_0(\tau) = e^T_1 X^{-1} Y \mathrm{e}^{D_S\tau}  Y^{-1} X e_1\,,
\end{equation}
which can be rewritten as a sum of complex exponential terms, with $\{\mu_j\}$ being the eigenvalues of $S$
as
\begin{equation}
P_0(\tau) = \sum_{j=1}^{d^2}\hat{z}_{1,j}z_{j,1} \mathrm{e}^{\mu_j\tau}\,.
\end{equation}
This expression corresponds to the analogue of Eq.~\eqref{eq:C2} in the main text. Since $S$ is by definition a \emph{Kronecker sum} of $Q^*$ and $Q$ with eigenvalues $\{q_j^*\}$ and $\{q_j\}$ respectively, the spectrum of $S$ consists of the sums $q_j^*+q_k$ with $j,k=1,\dots,d$. It is closed under complex conjugation (for each element of the set its complex conjugate is also element of the set), as well as the coefficient set $\{\hat{z}_{1,j}z_{j,1}\}$. This ensures that $P_0$ is real-valued. Being related to $Q$ (which consists of a skew-hermitian matrix (with imaginary spectrum) and negative definite matrices), we have that $\mathrm{Re}\,\mu_j<0$ for each $j$, such that all summands vanish sufficiently fast and $P_0$ is normalisable. Furthermore, the dominance of the damping factors over the oscillatory components ensures the positivity of $P_0$ (in particular, the $\mu_j$ with the least damping is always real-valued).
Analogously, for $P_1(\tau)$ we obtain
\begin{equation}
\sum_{j,k=1}^{d^2}\kern-.2em\hat{z}_{1,j}\mathcal{M}_{j,k}z_{k,1} \left((1-\delta_{j,k})\frac{\mathrm{e}^{\mu_k \tau} - \mathrm{e}^{\mu_j \tau}}{\mu_k-\mu_j}+\delta_{j,k}\tau \mathrm{e}^{\mu_j\tau}\right)
\end{equation}
with the Kronecker delta $\delta_{j,k}$,
\begin{equation}
	f(\tau)=\sum_{m,n} c_{m,n} \tau^m \mathrm{e}^{\mu_n \tau}. 
\end{equation}
Assuming that the terms $\tau^m \mathrm{e}^{\mu_n \tau}$ are linearly independent, in principle one can always single out these contributions as well as their corresponding prefactors $c_{m,n}$. This gives us the chance to extract the coefficients $\{\hat{z}_{1,j}z_{j,1}\}$ and the eigenvalues $\{\mu_i\}$ from $P_0$, provided that no coefficient is identical to zero. Rearranging the values $\{\mu_i\}$ to a diagonal matrix in Kronecker sum form results in the matrix $\mathcal{D}$. One should note, however, that efficient spectral recovery algorithms like the \emph{matrix pencil method} do not straightforwardly work for functions such as $P_n$, $n\geq 2$, where the exponential functions are multiplied with powers of $\tau$.

In order to reconstruct the elements of the matrix $\mathcal{M}$ together with the off-diagonal elements of $Z$, we use a gauge argument: All probability functions $P_n$ are invariant under scaling and permutation of the eigenvectors in the matrices $X$ and $Y$ (except for the eigenvector of $T$ corresponding to eigenvalue zero). This allows us to require all but one $\hat{z}_{1,j}$ to be equal to one, and immediately obtain the according number $z_{j,1}$. The remaining coefficient can then be determined via the normalisation constraint 
\be
\sum_{j=1}^{d^2}\hat{z}_{1,j}z_{j,1}=1,
\ee 
so that all $z_{j,1}$ are known. This can be used to obtain the diagonal elements $\mathcal{M}_{j,j}$ from $\mathcal{M}$. For the remaining matrix elements, only the symmetric elements $\mathcal{M}_{j,k}+\mathcal{M}_{k,j}$ (but not their constituents) are directly accessible since
\be
\label{eq:M sym}
&&\sum_{\substack{j,k=1\\
j\neq k
}
}^{d^{2}}\kern-0.2em \hat{z}_{1,j}\mathcal{M}_{j,k}z_{k,1}\frac{\mathrm{e}^{\mu_{k}\tau}-\mathrm{e}^{\mu_{j}\tau}}{\mu_{k}-\mu_{j}}=\nonumber\\
&& \sum_{j<k}^{d^{2}}\kern-0.2em (\hat{z}_{1,j}\mathcal{M}_{j,k}z_{k,1}+\hat{z}_{1,k}\mathcal{M}_{k,j}z_{j,1})\frac{\mathrm{e}^{\mu_{k}\tau}-\mathrm{e}^{\mu_{j}\tau}}{\mu_{k}-\mu_{j}}.
\ee
However, this does not constitute a limitation for the reconstruction of the matrices $Q$ and $R$ of the ancillary system. To this end, we make use of the inner structure of $\mathcal{M}$. The diagonal matrix $\mathcal{D}$ with eigenvalues $\mu_j$ can be reordered such that it has the form $\mathcal{D}=D_Q^* \otimes \one + \one \otimes D_Q$ with diagonal $D_Q$ consisting of the eigenvalues of $Q$. Reordering the eigenvectors in $Y$ accordingly, we can assume that the matrix $Y$ and hence the matrix $\mathcal{M}$ have the form of a Kronecker product
\be
\mathcal{M}=R_\mathrm{rec}^*\otimes R_\mathrm{rec}\,.
\ee
Here $R_\mathrm{rec}=(r_{j,k})\in\mathbb{C}^{d\times d}$ is in general not diagonal. The symmetrised components of $\mathcal{M}$ can then be written as $r_{j,k}^*r_{l,m}+r_{k,j}^*r_{m,l}$ and the constituents $r_j$ can be determined (up to a phase factor) by equating them with the coefficients in Eq.~\eqref{eq:M sym}. The according equation system can then be solved.

The important point is that $R_\mathrm{rec}$ and $Q_\mathrm{rec}:=D_Q$ are valid cMPS parameter matrices in the same gauge and hence are sufficient for reconstruction with the same argument as in Ref.\ \cite[III.E]{LongPaper}. Let us 
note that concrete values of the basis transformation matrices $X$ and $Y$ are in fact never used or needed in the reconstruction procedure. From $R_\mathrm{rec}$ and $Q_\mathrm{rec}$, we can compute all quantities we need to establish the correlation and counting probability functions, in particular $\mathcal{M}$ and $\mathcal{D}$. Regauging $R_\mathrm{rec}$ and $Q_\mathrm{rec}$ such that the orthonormalisation condition \cite{cMPS1} is fulfilled, yields a reconstruction of the free Hamiltonian $K_\mathrm{rec}$ of the ancillary system.



%

\end{document}